\documentclass[%
 aip,
 amsmath,amssymb,
preprint,%
]{revtex4-2}

\newcommand\beq{\begin{equation}}
\newcommand\eeq{\end{equation}}

\usepackage{graphicx}% Include figure files
\usepackage{dcolumn}% Align table columns on decimal point
\usepackage{bm}% bold math
\usepackage[utf8]{inputenc}
\usepackage[T1]{fontenc}
\usepackage{mathptmx}
\usepackage{etoolbox}
\usepackage[colorlinks, linkcolor= blue, citecolor = blue, urlcolor=blue]{hyperref}

\makeatletter
\def\@email#1#2{%
 \endgroup
 \patchcmd{\titleblock@produce}
  {\frontmatter@RRAPformat}
  {\frontmatter@RRAPformat{\produce@RRAP{*#1\href{mailto:#2}{#2}}}\frontmatter@RRAPformat}
  {}{}
}%
\makeatother

\begin{document}

\preprint{AIP/123-QED}

\title[]{Multiple actions of time-resolved short-pulsed metamaterials}
% Force line breaks with \\
\author{Giuseppe Castaldi}%
\affiliation{ 
	Fields \& Waves Lab, Department of Engineering, University of Sannio, I-82100, Benevento, Italy%\\This line break forced with \textbackslash\textbackslash
}%
\author{Carlo Rizza}
\affiliation{Department of Physical and Chemical Sciences, University of L'Aquila, I-67100 L'Aquila, Italy}
\author{Nader Engheta}
\affiliation{Department of Electrical and Systems Engineering, University of Pennsylvania, Philadelphia, PA, 19104, USA}
\author{Vincenzo Galdi}
 \email{vgaldi@unisannio.it}
\affiliation{ 
Fields \& Waves Lab, Department of Engineering, University of Sannio, I-82100, Benevento, Italy%\\This line break forced with \textbackslash\textbackslash
}%

\date{\today}% It is always \today, today,

%%%%%%%%%%%%%%%%%%%% Created: 		03/09/2022
%%%%%%%%%%%%%%%%%%%% Last revised:  16/12/2022

\begin{abstract}
Recently, it has been shown that temporal metamaterials based on impulsive modulations of the constitutive parameters (of duration much smaller than a characteristic electromagnetic timescale) may exhibit a nonlocal response that can be harnessed so as to perform elementary analog computing on an impinging wavepacket. These short-pulsed metamaterials can be viewed as the temporal analog of conventional (spatial) metasurfaces. Here, inspired by the analogy with cascaded metasurfaces, we leverage this concept and take it one step further, by showing that short-pulsed metamaterials can be utilized as elementary bricks for more complex computations. To this aim, we develop a simple, approximate approach to
systematically model the  multiple actions of time-resolved short-pulsed metamaterials. Via a number of representative examples, we illustrate the computational capabilities enabled by this approach, in terms of simple and composed operations, and validate it against a rigorous numerical solution. Our results indicate that the temporal dimension may provide new degrees of freedom and design approaches in the emerging field of computational metamaterials, in addition or as an alternative to conventional spatially variant platforms.
\end{abstract}

\maketitle
The long-standing research field of electromagnetic wave propagation in time-varying media\cite{Morgenthaler:1958vm,Oliner:1961wp,Felsen:1970wp,Fante:1971to} 
is currently witnessing a renewed surge of interest under the overarching framework of {\em temporal} and {\em space-time} metamaterials.\cite{Engheta:2021mw,Caloz:2020sm1,Caloz:2020sm2,Pacheco:2022tv} These emerging metastructures are characterized by time modulations of the constitutive parameters, instead of (or in addition to) the conventional spatial ones.
As also summarized in a recent review,\cite{Galiffi:2022po} a wide variety of temporal analogies for effects, concepts and tools that are well-established for conventional (spatially modulated) metamaterials have been explored, including temporal boundaries \cite{Xiao:2014ra} and slabs \cite{Mendonca:2003tb,Ramaccia:2020lp}, homogenization \cite{Pacheco:2020em,Huidobro:2021ht,Rizza:2022ne}, diffraction gratings \cite{Taravati:2019gs,Galiffi:2020wa}, 
photonic time crystals \cite{Romero:2016tp,Lyubarov:2022ae},
impedance transformers \cite{Shlivinski:2018bt,Pacheco:2020at,Castaldi:2021es,Galiffi:2022tp},  
filters \cite{Ramaccia:2021tm,Rizza:2022sp,Silbiger:2022of,Castaldi:2022he}, 
spectral causality \cite{Hayran:2021sc},
Faraday rotation \cite{Li:2022na}, 
and Brewster angle \cite{Pacheco:2021te}.  Within this framework, it is also worth highlighting that reliance on time-varying media may also enable overcoming certain fundamental limitations in (linear, time-invariant) electromagnetic systems.\cite{Shlivinski:2018bt,Li:2021ts,Hayran:2022cf}
Although it has been pointed out that temporal modulations of the constitutive parameters are subject to inherent dispersion- and energy-related constraints,\cite{Hayran:2022hb} which pose unique challenges to their practical implementation, experimental studies are gaining momentum, and recent results have demonstrated some of the above concepts.\cite{Moussa:2022oo,Wang:2022mb,Liu:2022pa}

Of particular interest for the present study are the analog signal-processing capabilities of temporal metamaterials that have recently been put forward, \cite{Rizza:2022ne,Rizza:2022sp,Silbiger:2022of,Galiffi:2022tp} along the lines of spatially variant  computational metamaterials.\cite{Silva:2014pm,Zangeneh-Nejad:2020ac} In particular, we recently introduced the concept of {\em short-pulsed} metamaterials (SPMs),\cite{Rizza:2022sp} entailing temporal modulations of the dielectric permittivity of duration much smaller than the characteristic wave-dynamical timescale; in the space-time analogy, these can be viewed as the temporal counterpart of metasurfaces.
By suitably tailoring the parameters, we showed that SPMs can perform elementary analog computing, such as first and second derivatives, on an impinging wavepacket.\cite{Rizza:2022sp} 

Here, we further expand this idea, by demonstrating that SPMs can be utilized as elementary bricks for more complex computations. To this aim, inspired by the spatial analogy with cascaded metasurfaces,\cite{Raeker:2019cm} we study the multiple actions of time-resolved SPMs.

%############################################################
%                Figure1
%
\begin{figure}
	\centering
	\includegraphics[width=\linewidth]{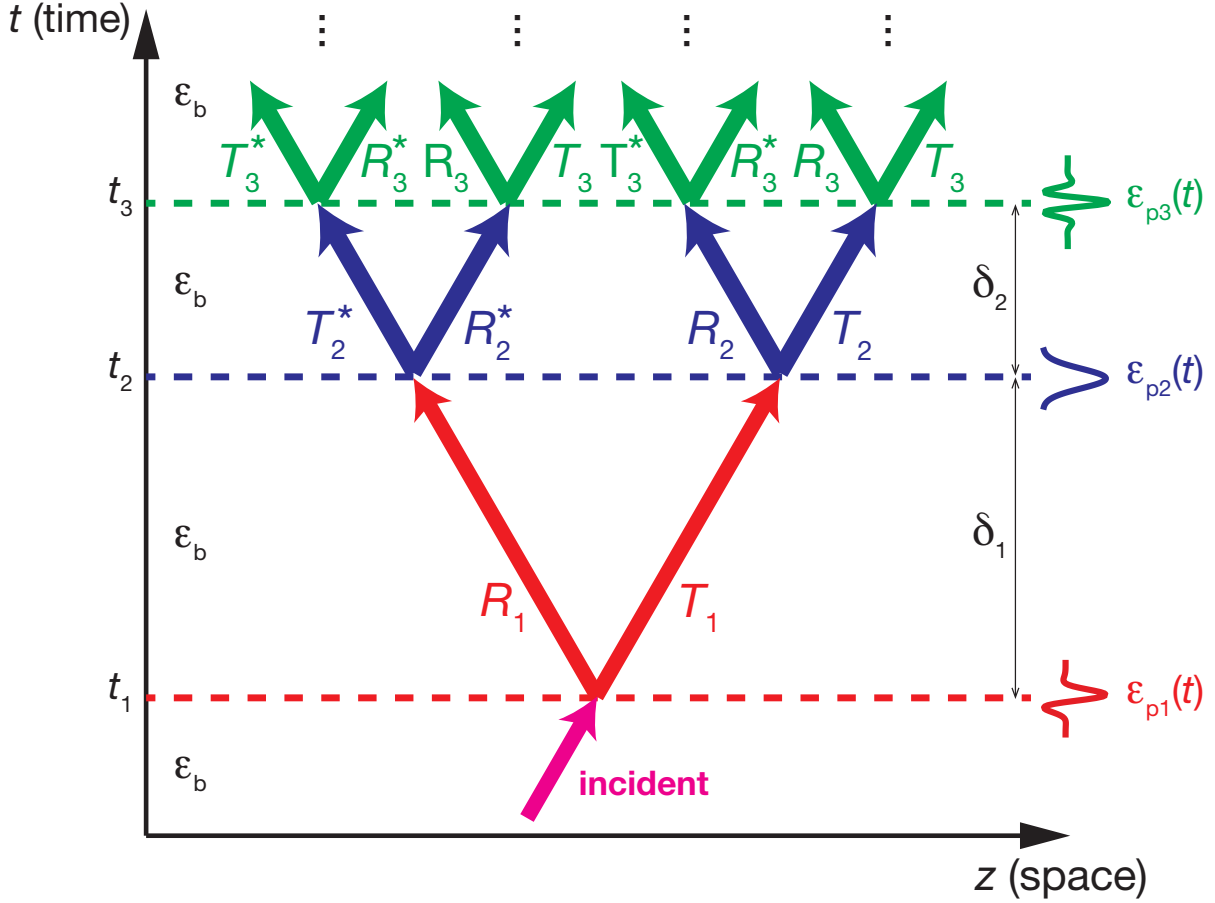}
	\caption{Problem schematic (details in the text). Note the phase conjugation of the reflection and transmission coefficients for backward incidence.}
	\label{Figure1}
\end{figure}
%############################################################

Referring to Ref. \citenum{Rizza:2022sp} for details, we model a generic SPM as 
a relative-permittivity waveform $\varepsilon_p\left(t\right)$ of duration $\tau$, starting at time $t=t_0$, in a stationary  background with relative permittivity
$\varepsilon_b$; all media are assumed as non-magnetic. Assuming as a reference temporal scale the characteristic duration $\Delta t$ of a wavepacket, the SPM regime of interest is characterized by $\tau\ll\Delta t$.
Via a multiscale approach, we showed that, in this asymptotic limit, the SPM response can be approximated in terms of a {\em nonlocal} temporal boundary,
described by a minimal number of parameters.\cite{Rizza:2022sp} In particular, by assuming plane-wave incidence with electric-induction
$D\left(z,t\right)=\mbox{Re}\left[d\left(k,t\right)\exp\left(ikz\right)\right]$,  with $k$ denoting the conserved wavenumber, the 
 reflection (i.e., backward-wave) and transmission (i.e., forward-wave) coefficients can be approximated as\cite{Rizza:2022sp}
\begin{subequations}
	\begin{eqnarray}
	\!\!\!\!\!\!\!\!\!{R}\left(\kappa\right)&=&\frac{d_{bw}\left(k,t_0+\tau\right)}{d_{in}\left(k,t_0\right)}
	\approx \left[i \pi \left( \frac{n_b}{\varepsilon_{eff}} - \frac{1}{n_b} \right) \kappa  +2\pi \beta_0 \kappa^2\right], 
		\label{eq:R}\\
		\!\!\!\!\!\!\!{T}\!\!\left(\kappa\right)&=&\frac{d_{fw}\left(k,t_0+\tau\right)}{d_{in}\left(k,t_0\right)}
		\approx \left[1\!-\! i\pi\! \left( \frac{n_b}{\varepsilon_{eff}} \!+\! \frac{1}{n_b} \right) \!\!\kappa  \!-\!\frac{2\pi^2}{\varepsilon_{eff}} 
		\!\kappa^2\right], 
		\label{eq:T}
	\end{eqnarray}
\label{eq:RT}
\end{subequations}
where the subscripts ``in'', ``fw'' and ``bw'' indicate the incident, forward and backward terms, respectively, and the explicit dependence on the normalized wavenumber $\kappa=k c n_b \tau/(2 \pi)$ highlights the anticipated nonlocal character. Moreover,  $\omega=ck$, $c$ and $n_b=\sqrt{\varepsilon_b}$ denote the background-medium angular frequency, wavespeed, and refractive index, respectively, and
\beq
\varepsilon_{eff}=a_0^{-1}, \quad \beta_0=2 \mbox{Im} \left(\sum_{n=1}^{+\infty} \frac{a_n}{n}\right),
\eeq
are the effective relative permittivity and a symmetry parameter, respectively, describing the SPM waveform, which can be computed from the Fourier coefficients $a_n$ of $\varepsilon_p^{-1}\left(t\right)$ (see Ref. \citenum{Rizza:2022sp} for details). 
 The expressions in Eqs. (\ref{eq:RT}) neglect terms of order ${\cal O}(\kappa^3)$ and higher, and have been validated against exact (temporal slab\cite{Mendonca:2003tb}) and rigorous numerical solutions.\cite{Rizza:2022sp} In the weak-dispersion regime $\kappa\ll1$ of interest here, it is evident that the transmission coefficient in Eq. (\ref{eq:T}) is dominated by the local response ($T\approx 1$). Conversely, the nonlocal terms are dominant in the reflection coefficient in Eq. (\ref{eq:R}), and their amplitudes can be tailored so as to perform a  first derivative (for $\varepsilon_{eff}\ne \varepsilon_b$, $\beta_0=0$, i.e., $R\propto i\kappa$), a second derivative (for $\varepsilon_{eff}=\varepsilon_b$, $\beta_0\ne0$, i.e., $R\propto \kappa^2$), or a combination of the two.

Referring to Fig. \ref{Figure1} for a conceptual illustration, in this study, we consider a series of SPMs, characterized by relative-permittivity waveforms $\varepsilon_{pm}(t)$  of duration $\tau_m\ll \Delta t$ ($m=1,2,...,M$) in a stationary background. The starting instants $t_m$ are chosen so as to ensure that these SPMs are well resolved in time, i.e., $\tau_m\ll \delta_m=t_{m+1}-t_m$.
In order to obtain a simple, insightful model of the compound response, we represent each of the SPMs as a nonlocal temporal boundary, with reflection and transmission coefficients $R_m$ and $T_m$ as in Eqs. (\ref{eq:RT}), and study their interactions in a space-time diagram, as for conventional (local) temporal boundaries.\cite{Mai:2022fa}  In essence, as schematically illustrated in Fig. \ref{Figure1}, at each temporal boundary, the wave is split into a forward and backward component, and these ramifications generate a tree diagram. Therefore, the response of a number $M$ of SPMs  can be expressed as a sum of $2^M$ components, whose complex amplitudes can be obtained by following the corresponding tree branches, and accounting for the appropriate reflection or transmission coefficients at each boundary. Note that the reflection and transmission coefficients in Eq. (\ref{eq:RT}) are calculated assuming forward incidence,\cite{Rizza:2022sp} and hence their conjugate (time-reversed) version must be considered when encountering backward-incidence (see Fig. \ref{Figure1}). Likewise, the boundary-to-boundary travel times are accounted for via phase-shift factors $\exp(\pm i\omega \delta_m)$ (for backward and forward propagation, respectively). 
Thus, for instance, for $M=2$ (i.e., two SPMs), we obtain for the compound reflection and transmission coefficients
\begin{subequations}
	\begin{eqnarray}
		{R}_c &=&\frac{d_{bw}\left(k,t_2+\tau_2\right)}{d_{in}\left(k,t_1\right)}
		\approx R_1 T_2^*  \exp\left(i\omega\delta_1\right)+T_1R_2 \exp\left(-i\omega\delta_1\right),\\
		{T}_c &=&\frac{d_{fw}\left(k,t_2+\tau_2\right)}{d_{in}\left(k,t_1\right)}
		\approx T_1 T_2 \exp\left(-i\omega\delta_1\right)+R_1R_2^* \exp\left(i\omega\delta_1\right).
	\end{eqnarray}
\label{eq:RTc}
\end{subequations}
From an operator viewpoint, recalling that $T_{1,2}\approx1$, we observe that the reflection (backward) response contains a superposition of the (time-shifted) single operations  performed by each SPM, whereas the transmission (forward) response, besides a dominant local term, contains a (time-shifted) product of the two reflection coefficients, i.e., the {\em composition} of the two operations. For example, if the parameters of each SPM are tailored so as to perform the first derivative (i.e., $R_{1,2}\propto i\kappa$), the compound reflection response will contain two time-shifted first derivatives (i.e., $R_1$ and $R_2$), whereas the compound transmission response will contain a second derivative (i.e., $R_1R_2^*\propto \kappa^2$), besides an almost identical copy of the impinging wavepacket.
The above reasoning can be extended to an arbitrary number of SPMs, and symbolic manipulation tools can be exploited to deal with the growing number of terms. For instance, the response for the case $M=3$ is given in Appendix \ref{sec:AppA}.

%############################################################
%                Figure2
%
\begin{figure*}
	\centering
	\includegraphics[width=\linewidth]{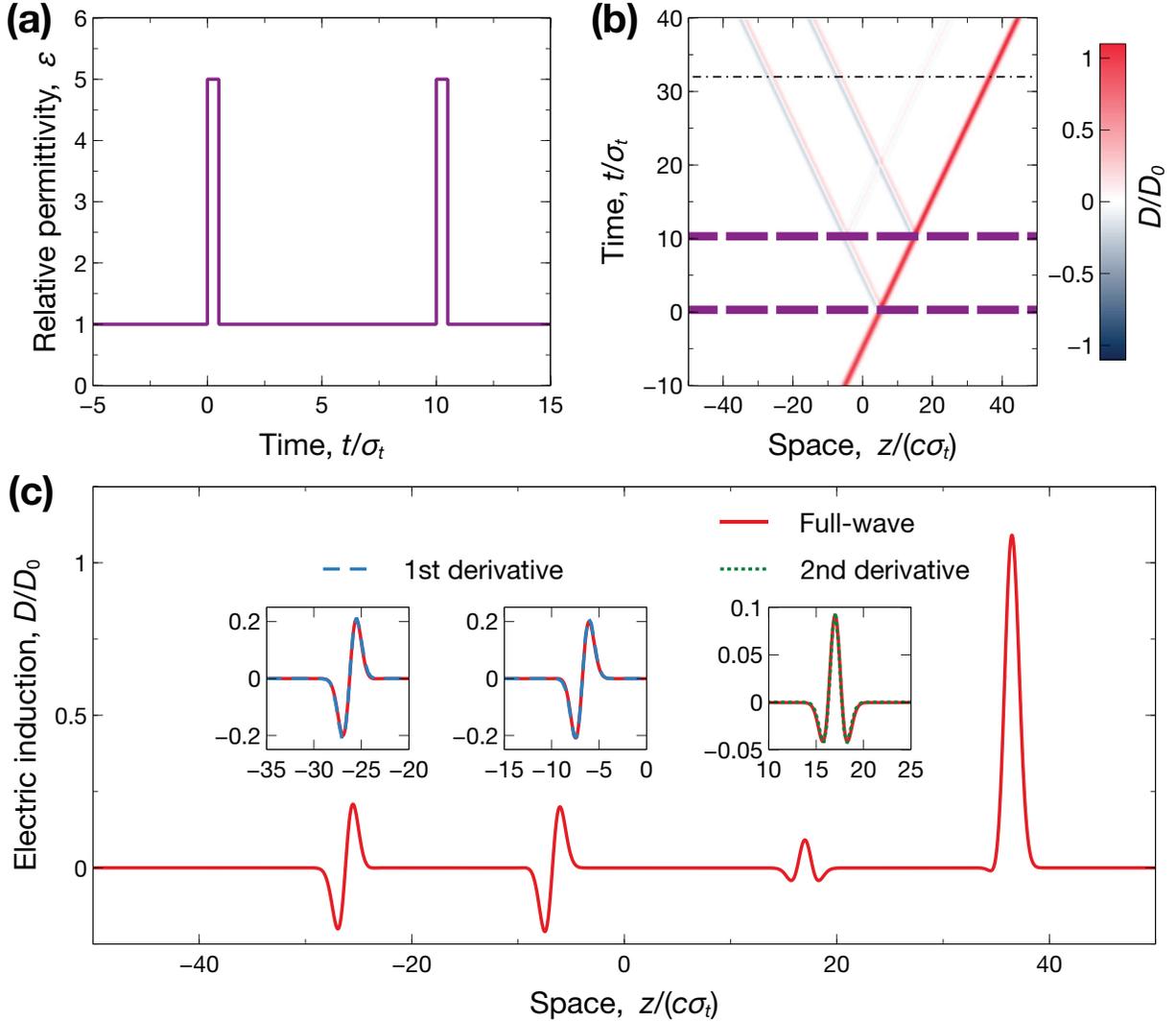}
	\caption{(a) Relative-permittivity waveform featuring two identical rectangular SPMs (temporal slabs), with duration $\tau_{1,2}=0.5\sigma_t$, amplitude $\varepsilon_{p1,2}=5$, and starting times $t_1=0$ and $t_2=10\sigma_t$, in a background medium with $\varepsilon_b=1$. (b) Space-time map [normalized electric induction, computed from Eqs. (\ref{eq:RTc})]. The thick purple-dashed lines indicate the temporal boundaries. (c) Spatial cut at $t=32\sigma_t$ [dash-dotted line in panel (b)], computed via a rigorous numerical solution. The three insets show some magnified views comparing the actual waveforms with the expected first and second derivatives.}
	\label{Figure2}
\end{figure*}
%############################################################

For illustration, we assume an impinging Gaussian wavepacket with profile (for $t<0$)
\beq
	D_{in}(z,t)=D_0 \exp\left\{-\left[\frac{z-c \left(t - t_s \right)}{c\sigma_t} \right]^2\right\},
	\label{eq:gauss}
\eeq 
where $D_0$ is a constant amplitude, $\sigma_t$ a characteristic timescale, and $t_s=-5 \sigma_t$.
For validation, we utilize a reference solution obtained via the rigorous numerical approach detailed in Refs. \citenum{Rizza:2022ne,Rizza:2022sp} (not repeated here for brevity).

As a first illustrative example, as shown in Fig. \ref{Figure2}a, we consider a scenario featuring two identical rectangular SPMs (temporal slabs, with $\tau_{1,2}=0.5\sigma_t$ in order to fulfill the SPM condition), each performing a first derivative.\cite{Rizza:2022sp} Figure \ref{Figure2}b shows a space-time map, computed from Eqs. (\ref{eq:RTc}), from which we observe that the phenomenon qualitatively resembles the schematic description in Fig. \ref{Figure1}. For a more quantitative assessment, Fig. \ref{Figure2}c shows a spatial cut (at a fixed time chosen so as all contributions are well resolved) numerically computed via the rigorous reference solution,\cite{Rizza:2022sp} with the three insets comparing the actual waveforms with the expected (first or second) derivatives. As we can observe, the agreement is excellent.

\begin{figure*}
	\centering
	\includegraphics[width=\linewidth]{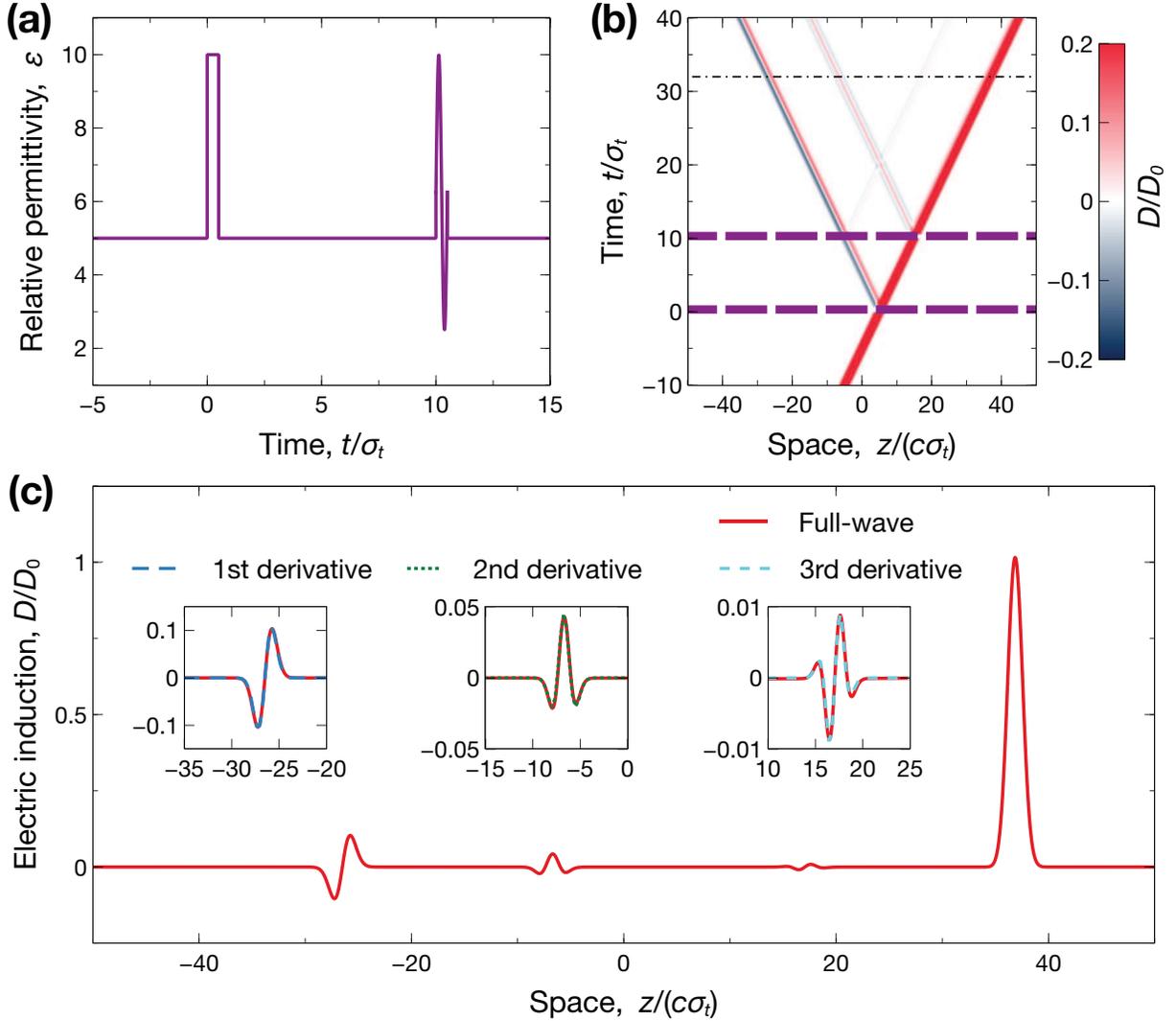}
	\caption{(a) Relative-permittivity waveform featuring a rectangular SPM (temporal slab, with $\tau_1=0.5\sigma_t$, $\varepsilon_{p1}=10$, $t_1=0$) and a sinusoidal one [Eq. (\ref{eq:ep2}), with $\varepsilon_{m2}=6.25$, $\tau_2=0.5\sigma_t$, $t_2=10\sigma_t$, $\Delta=0.6$, $\phi=-\pi/2$], in a background medium with $\varepsilon_b=5$. (b) Space-time map [normalized electric induction, computed from Eqs. (\ref{eq:RTc})]. The thick purple-dashed lines indicate the temporal boundaries, and the colorscale is suitably saturated so as to show the weakest waveform. (c) Spatial cut at $t=68\sigma_t$ [dash-dotted line in panel (b)], computed via a rigorous numerical solution. The three insets show some magnified views comparing the actual waveforms with the expected first, second, and third derivatives.}
	\label{Figure3}
\end{figure*}

As a second example, as shown in Fig. \ref{Figure3}a, we consider a more complex scenario, featuring a rectangular SPM (performing a first derivative) paired with a sinusoidal one with a relative-permittivity waveform (for $t_2<t<t_2+\tau_2$)
\beq
\varepsilon_{p2}\left(t\right)=\varepsilon_m\left\{1+\Delta\cos\left[2\pi\left(\frac{t-t_2}{\tau_2}\right)+\phi\right]\right\},
\label{eq:ep2}
\eeq
and parameters chosen so as to perform a second derivative.\cite{Rizza:2022sp} In this case, from Eqs. (\ref{eq:RTc}), we expect a first and second derivative in the backward waveforms, and a third derivative (in addition to the local term) in the forward ones.
Figure \ref{Figure3}b shows the resulting space-time map, with the colorscale suitably saturated to display all contributions. From the spatial cut (at $t=32\sigma_t$) shown in Fig. \ref{Figure3}c we observe, once again, an excellent agreement with the expected results. We note, however, that the amplitudes of the high-order derivatives decreases rapidly, and (as anticipated from the space-time map) the third-derivative term is rather weak. 
This decreasing efficiency is somehow inherent of the underlying physics, based on multiple interactions. Nevertheless, we remark that the parameters in the above examples were  chosen to illustrate the basic phenomenology, rather than to maximize any specific effect.
Therefore, there is room for moderate improvements, also taking into account that our time-varying platform is inherently active, and as such not bound by power conservation for electromagnetic signals.

\begin{figure*}
	\centering
	\includegraphics[width=\linewidth]{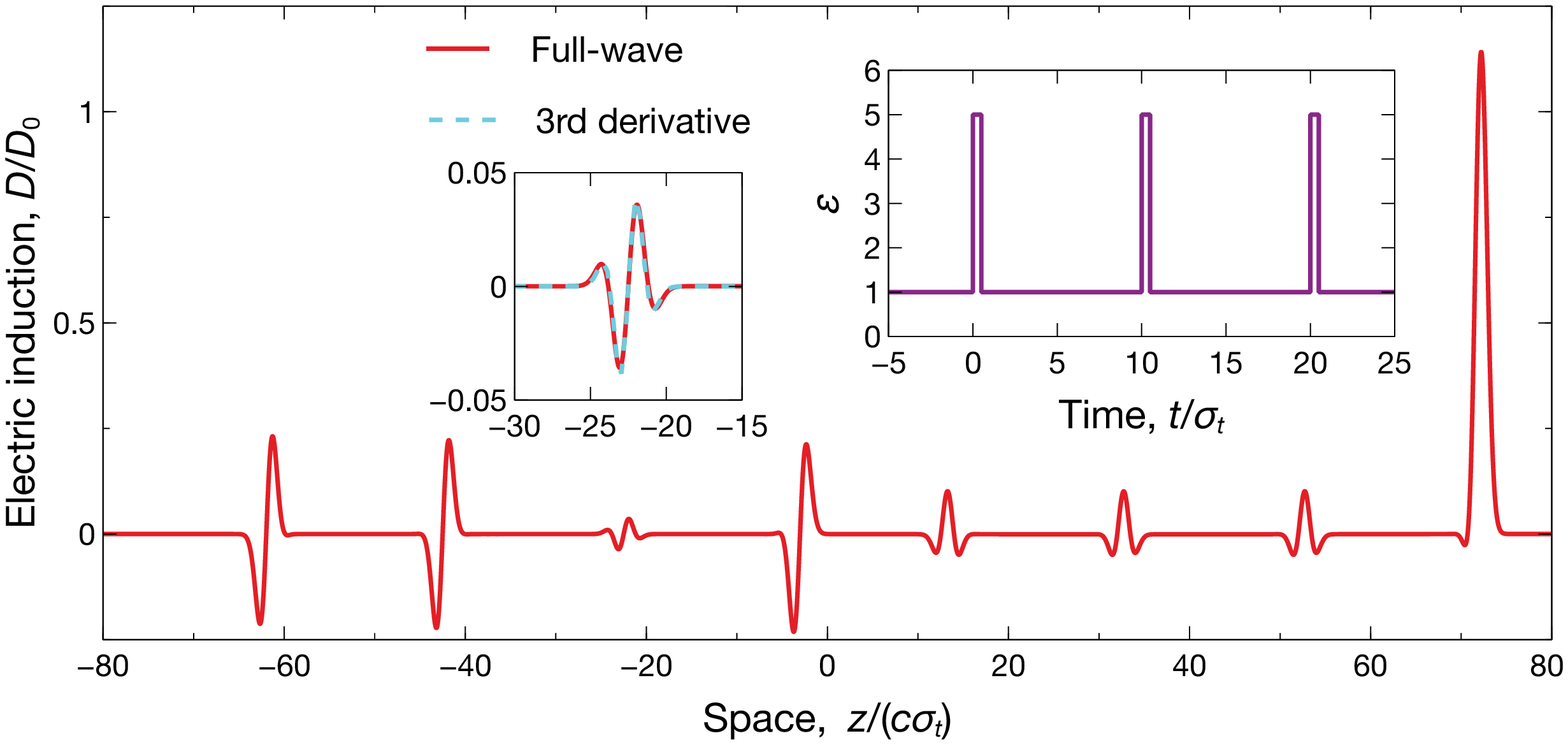}
	\caption{Spatial cut at $t=74\sigma_t$ of the normalized electric induction, computed via a rigorous numerical solution, pertaining to a scenario featuring three identical rectangular SPMs (temporal slabs, with $\tau_{1,2,3}=0.5\sigma_t$, $\varepsilon_{p1,2,3}=5$, $t_1=0$, $t_2=10\sigma_t$, $t_3=20\sigma_t$, in a background medium with $\varepsilon_b=1$ (see right inset). The left inset shows a magnified view comparing the actual waveform with the expected third derivative.}
	\label{Figure4}
\end{figure*}

For instance, Fig. \ref{Figure4} shows the results pertaining to a scenario featuring three identical rectangular SPMs (temporal slabs), for which, among other terms, a third derivative is expected in the backward waveforms [see Eqs. (\ref{eq:RTc3}) in Appendix \ref{sec:AppA}]. 
Focusing our attention on this term, once again, we observe an excellent agreement with the theoretical prediction, with an amplitude that is larger by a factor $\sim 5$ than the previous example in Fig. \ref{Figure3}. The comparisons for the other derivative terms (not shown for brevity), are in line with what observed in the previous examples.

%############################################################
%                Figure3
%
%############################################################

%############################################################
%                Figure4
%
%############################################################

To sum up, we have extended the concept of SPM to a more general, multiple-interaction scenario. Specifically, we have considered systems of multiple, time-resolved SPMs, each tailored so as to perform an elementary analog computation, and have studied the compound reflection and transmission responses, illustrating the more complex, composed operations that can be attained. Our results, validated against rigorous numerical simulations, illustrate the potential capabilities of SPMs to serve as elementary bricks in more complex analog-computing systems. In particular, our proposed approach endows more flexibility and degrees of freedom for the synthesis of operations (e.g., higher-order derivatives) that would be impossible or impractical to realize with a single SPM, and for their placement either in the reflection (backward) or transmission (forward) responses.

As for the practical feasibility, we acknowledge that deep, short-pulsed modulations of the constitutive parameters remain very challenging from the technological viewpoint, especially at high frequencies.\cite{Hayran:2022hb} For instance, at microwave frequencies, the implementation could rely on transmission-line metamaterials with rapidly switchable capacitance, which have recently been demonstrated to induce temporal boundaries.\cite{Moussa:2022oo} At terahertz frequencies, reliance could be made on infrared femtosecond laser pulses, which can induce a temporal modulation of the dielectric permittivity of a semiconductor (e.g., GaAs, Si) with significant depth, on a timescale of $\tau\sim$100 fs.\cite{Kamaraju:2014sc,Yang:2017tg} In principle, this platform could enable the implementation of SPMs  for manipulating terahertz wavepackets with characteristic timescales $\Delta t\sim$1 ps. 

Current and future studies are aimed at developing more general synthesis approaches, as well as further extending the SPM concept to more complex scenarios involving anisotropy and/or spatio-temporal modulations.

\section*{Acknowledgment}
G. C. and V. G.  acknowledge partial support from the University of Sannio via the FRA 2021 Program.
N. E. acknowledges partial support from the Simons Foundation/Collaboration on Symmetry-Driven Extreme Wave Phenomena grant \#733684.

\section*{Author Declarations}

\subsection*{Conflict of interest}
The authors have no conflicts to disclose.

\subsection*{Data Availability Statement}
The data that support the findings of this study are available from the corresponding author upon reasonable request.

\appendix

\section{Compound reflection and transmission coefficients for $M=3$}
\label{sec:AppA}
For the case $M=3$, we have a grand total of eight contributions (four forward and four backward). Similar to the case $M=2$, by following the various branches of the tree-diagram in Fig. \ref{Figure1}, we obtain
%\begin{widetext}
\begin{subequations}
	\begin{eqnarray}
		R_c =\frac{d_{bw}\left(k,t_3+\tau_3\right)}{d_{in}\left(k,t_1\right)}
		&\approx& R_1 T_2^* T_3^*  \exp\left[i\omega\left(\delta_1+\delta_2\right)\right]\nonumber\\
		&+&R_1 R_2^* R_3  \exp\left[i\omega\left(\delta_1-\delta_2\right)\right]\nonumber\\
		&+&T_1 R_2 T_3^* \exp\left[-i\omega\left(\delta_1-\delta_2\right)\right]\nonumber\\
		&+&T_1 T_2 R_3  \exp\left[-i\omega\left(\delta_1+\delta_2\right)\right],\\
			T_c =\frac{d_{fw}\left(k,t_3+\tau_3\right)}{d_{in}\left(k,t_1\right)}
			&\approx& T_1 T_2 T_3  \exp\left[-i\omega\left(\delta_1+\delta_2\right)\right]\nonumber\\
		&+&T_1 R_2 R_3^*  \exp\left[-i\omega\left(\delta_1-\delta_2\right)\right]\nonumber\\
		&+&R_1 R_2^* T_3 \exp\left[i\omega\left(\delta_1-\delta_2\right)\right]\nonumber\\
		&+&R_1 T_2^* R_3^*  \exp\left[i\omega\left(\delta_1+\delta_2\right)\right].
	\end{eqnarray}
\label{eq:RTc3}
\end{subequations}	
%	\end{widetext}
Among the various terms, we identify some representing single operations, as well as the composition of two or three operations. For instance, if each of the SPMs is tailored so as to perform the first derivative, the backward response will contain three (time-shifted) first derivatives and a third derivative, whereas the forward response will contain three (time-shifted) second derivatives and an almost identical copy of the impinging wavepacket (see, e.g., Fig. \ref{Figure4}).

%\nocite{*}
%\bibliography{M-SPM}% Produces the bibliography via BibTeX.

%aipnum4-2.bst 2019-01-14 (MD) hand-edited version of apsrev4-1.bst

\end{document}